\documentclass[12pt,letterpaper]{article}
\usepackage{epsfig}
\usepackage{amsmath}
\usepackage[psamsfonts]{amssymb}
\usepackage{amsthm}
\usepackage{fullpage}
\usepackage{cite}
\usepackage{indentfirst}

\date{\today}
\def\be{\begin{equation}}
\def\ee{\end{equation}}
\def\bear{\begin{eqnarray}}
\def\eear{\end{eqnarray}}

\def\half{{ \frac{1}{2} }}

\def\D{{\overline D}}                              
\def\bv{{\bf v}}                               

\def\C{{{\bf C}^3}}                               
\def\a{{\alpha}}
\def\b{{\beta}}

\def\tht{{\theta}}
\def\lam{{\lambda}}

\begin{document}
%
\begin{titlepage}
\titlepage
\rightline{hep-th/0404246}
\rightline{ HUTP-04/A018}
\rightline{\today}
\vskip 1cm
\centerline{{\Huge  D-branes as Defects in the Calabi-Yau Crystal}}
\vskip 4cm
\centerline{  Natalia Saulina and 
Cumrun Vafa  }
\vskip 1cm
\begin{center}
 Jefferson Physical Laboratory
 \\  Harvard University \\
Cambridge, MA 02138, USA\\
\end{center}
\vskip 2cm
\abstract{We define the notion of A-model Lagrangian D-branes as 
introducing defects in the Calabi-Yau crystal.  The crystal melting in the
presence of these defects reproduces all genus string amplitudes as well
as leads to additional non-perturbative terms.
}
\end{titlepage}
\tableofcontents
\section{Introduction}
In \cite{ORV} a duality between topological A-type
string theory and statistical model of a three dimensional  melting crystal 
was proposed. In particular it was demonstrated that partition 
function $Z_{crystal}$ of the melting crystal corner computes
the closed string partition function $Z_{closed}$ in the $\C$ background.  
More precisely,
\be
Z_{crystal}(q)=Z_{closed}(q),\quad q=exp(-g_s)
\label{dual}
\ee
where  the string coupling 
constant $g_s$ 
is identified with the inverse of the temperature of the crystal.
Also in \cite{ORV}  it was explained how to  compute  the 
topological vertex \cite{AKMV} in
the crystal picture.
In \cite{INOV} this duality was interpreted
as coming from quantum gravitational foam for
 target description of the A-model strings, which is 
 a ``quantum K\"ahler gravity''.  
Each geometry contributing to the quantum foam corresponds 
to a configuration of molten crystal (roughly speaking
each atom removal from the crystal changes the quantum foam by blowing 
up of a corresponding point).
It was proposed in \cite{INOV}
that in the limit $g_s \ll 1$ the quantum foam (the melting crystal )
 is a fundamental
description of  physics at short distances $\sim g_s^{\half} l_s.$ 
String
 emerges in the high temperature
limit ($1/T=g_s<<1$) as an effective description of the theory at
the string scale $l_s$.  A related mathematical proposal
was also presented in \cite{OPN}.
Further developments have shed light on this duality:
It was conjectured in\cite{NV} that 
 A and B models on the same Calabi-Yau are S-dual to one another.
Moreover it was shown in \cite{NOV} that this S-duality follows
from type IIB S-duality and that the atoms of the crystal
can also be viewed as D(-1) instantons of the B-model.
See also \cite{Kap}.

In this note we continue to investigate the proposed duality
and would like to give the statistical interpretation for
non-compact Lagrangian D-branes in the context of $\C.$ It turns out that the
statistical interpretation of Lagrangian D-branes is introducing
defects to the crystal.  For the case of branes at some
particular framing we find that the topological string amplitudes
are identical to that of melting crystal in the presence of the
defects introduced by D-branes.  We also find the defect interpretation
of the lattice for arbitrary framing and also in the presence
of anti-D-branes.  However for these cases the corresponding
melting rule needs to be modified, similar to the melting
rule in the context of local toric 3-folds (when we have more
than one corner for the lattice).  We will make some comments
about what this melting rule would look like for these
more general branes, but will not develop
it in full detail in this paper.

The organization of this paper is as follows:
In section 2 we review the results of \cite{ORV}. In section 3 we 
present the geometry of modified crystal for branes and anti-branes
and also give
melting crystal picture for certain configurations of
D-branes and anti-D-branes.
In section 4 we discuss general brane and anti-brane configurations
at arbitrary framing.
In section 5 we show that our melting crystal picture of the D-branes
is compatible with the semiclassical limit of
small string coupling.
 
\section{ Review of the closed topological strings/melting crystal duality}
\begin{figure}
\begin{center}
\epsfig{file=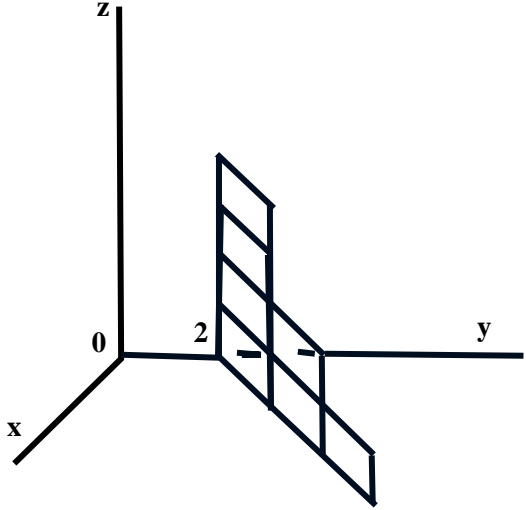,width=70mm}
\end{center}
\caption{ We draw 2d partition $\mu$ in a diagonal slice at $t=-2.$
The numbers of rows are  $\mu_1=4,\quad \mu_2=2,\quad \mu_3=1$. The fermion state is
$\vert \mu>=\psi^*_{-7/2}\psi_{-5/2}\psi^*_{-1/2}\psi_{-1/2}\vert 0>.$}
\label{fig1}
\end{figure}

In this section we review the duality between  closed A-model topological
string theory on $\C$ and statistical model of melting crystal corner
proposed in \cite{ORV}.
The A-model topological string theory studies
holomorphic maps from Riemann surfaces to the K\" ahler target space. 
\be
\Phi: \Sigma \rightarrow X
\label{map}
\ee
The partition function of this theory is given by
\be
Z_A=exp(\sum g_s^{2g-2}{\cal F}_g ),\quad
{\cal F}_g=\int_{{\cal M}_{maps,g}} e^{-A_C}
\label{topa}
\ee
where $A_C$ is the area of the holomorphic curve $C=\Phi(\Sigma_g)$ of genus
$g$,
and integration
goes over the moduli space of maps ${\cal M}_{maps,g}$ with a suitable measure.
Here $g_s$ denotes the string coupling constant.
When the  volume of $X$ is large the leading contribution to $Z_A$
comes from constant maps and the moduli space becomes
a product ${\cal M}_{maps,g}={\cal M}_g\times X$, where
${\cal M}_g$ is the moduli space of genus $g$ curves.
In this large volume limit the expressions for ${\cal F}_g$ were obtained in \cite{GV,FP}. 
As follows from the results of  \cite{GV,FP} the closed string partition function on $\C$ 
(taking its Euler character to be $2$) is given by the
McMahon function:
\be
Z_{closed}(q)=M(q):=\prod_{n=1}^{\infty}(1-q^n)^{-n}, \quad q=e^{-g_s}
\label{zclosed}
\ee
Now let us recall how the same function 
arises in the dual melting crystal picture \cite{ORV}.
Consider cubic lattice in the positive octant of $R^3$ with
lattice spacing given by $g_s$ and define the rules of melting.
One can remove atom with coordinates $(x_0,y_0,z_0)$ from the lattice
if and only if all the following sites are already vacant:
\be
(x,y_0,z_0) \quad x< x_0, \quad (x_0,y,z_0) \quad y< y_0, \quad (x_0,y_0,z) \quad z< z_0
\label{rules}
\ee
With these rules, each configuration of $k$ molten atoms corresponds
to a ``3d Young diagram'' (a 3d partition) of $k$ boxes in the 
positive octant of $R^3.$ 
Now we  sum over the random  3d partitions
with the weight $q^{\# boxes},$ where $q=e^{-g_s}$ and
$g_s$ stands for inverse temperature of the crystal (measured
in units of chemical potential for removing atoms).
To compute 
\be
Z_{crystal}(q)=\sum_{3d \ partition \ \pi }q^{\# boxes}
\label{zcrystal}
\ee 
we recall that, as discussed in \cite{OR}, a 3-dimensional
partition $\pi$ can be thought as a sequence of 2d partitions
$\{ \mu(t)\},t\in {\bf Z}$ obtained from diagonal slicing of 
$\pi$ by planes $x-y=t.$
These 2d partitions obey the interlacing condition:
\be
\mu(t)<\mu(t+1),\quad t<0,\quad \quad \mu(t+1)<\mu(t) \quad t\ge 0
\label{inter}
\ee
where the  2d partitions $\mu$ and $\nu$ interlace,
and we write $\mu > \nu,$
if
\be
\mu_1\ge \nu_1\ge \mu_2\ge \nu_2\ldots
\label{interlacing}
\ee
Here and below we let $\mu_k$  denote the number
of squares in the  k-th row of $\mu.$
The diagonal slicing allows us to use a well known map
from 2d partitions to states in the NS sector of a complex fermion:
\be
\psi(z)=\sum_{n\in {\bf Z}} \psi_{n+\half} z^{-n-1},
\quad
\psi^*(z)=\sum_{n\in {\bf Z}} \psi^*_{n+\half} z^{-n-1},\quad 
\{ \psi_{n+\half},\psi^*_{-(m+\half)} \}=
\delta_{m,n}
\label{psi}
\ee
The state corresponding to a partition $\mu$ is given by
\be
\vert \mu \bigr >=\prod_{i=1}^d \psi^*_{-a_i}\psi_{-b_i}\vert 0\bigr >
\label{state}
\ee
where $d$ is the number of squares on the diagonal of $\mu$
and, denoting 
the transposed 2d partition  by $\mu^T,$ we define
\be
a_i=\mu_i-i+\half, \quad \quad b_i=\mu^T_i-i+\half
\label{ab}
\ee
The example of the  state/2d partition  correspondence is shown in
Figure \ref{fig1}.
 
Now, $Z_{crystal}(q)$ can be easily recast 
 in the operator language:
\be
Z_{crystal}(q)=\Biggl <0 \Biggl \vert \Bigl(\prod_{t=0}^{\infty} q^{L_0}\Gamma_{+}(1)\Bigr) q^{L_0}
\Bigl(\prod_{t=-\infty}^{-1}\Gamma_{-}(1)q^{L_0} \Bigr) \Biggr \vert 0\Biggr >
\label{oper}
\ee
Here $\vert 0>$ is the state annihilated by
all positive modes of $\psi$ and $\psi^*$ and operators $L_0,\Gamma_{\pm}(1)$
act in the following way.
\be 
L_0 \vert \mu>=\vert \mu \vert \vert \mu >
\label{lzero}
\ee
\be
\Gamma_{-}(1) \vert \mu>=\sum_{\nu > \mu}\vert \nu>,\quad
 \Gamma_{+}(1) \vert \mu>=\sum_{\nu < \mu}\vert \nu>
\label{gammas}
\ee
where $\vert \mu \vert$
stands for the number of squares in 2d partition $\mu.$
The final step in computing $Z_{crystal}(q)$ is to rewrite 
the correlator (\ref{oper})  in the form
\be
Z_{crystal}(q)= <0\vert \prod_{n=1}^{\infty}\Gamma_{+}\bigl (q^{n-\half}\bigr ) 
\prod_{m=1}^{\infty}\Gamma_{-}\bigl ( q^{-m+\half}\bigr )  \vert 0>
\label{operii}
\ee
where 
\be
\Gamma_{+}\bigl (q^{n-\half}\bigr )= q^{-(n-\half)L_0} \Gamma_{+}(1) q^{(n-\half)L_0},
\quad \Gamma_{-}\bigl (q^{-n+\half}\bigr )= q^{-(-n+\half)L_0} \Gamma_{-}(1) q^{(-n+\half)L_0}
\label{newoper}
\ee
We will say more about the operators $\Gamma_{\pm}(z)$ and their
simple realization in the bosonized picture in the next section.
Here we only need their  commutation relation
\be
\Gamma_{+}(z)\Gamma_{-}(z')=\bigl(1-z/z'\bigr)^{-1}\Gamma_{-}(z')\Gamma_{+}(z)
\label{gamcom}
\ee
which leads  immediately to :
\be
Z_{crystal}(q)=\prod_{n=1}^{\infty}(1-q^n)^{-n},\quad q=e^{-g_s}
\label{fin}
\ee
Comparing (\ref{zclosed}) with (\ref{fin}) finishes the proof
of the duality (\ref{dual}) for closed strings.
In \cite{ORV} it was shown how one can formulate the topological vertex
amplitudes \cite{AKMV} in terms of crystal melting using the above operator
formalism.
\section{String/Crystal duality with D-branes }
In this section we give the melting crystal interpretation of  D-branes
in topological A-type string theory on $\C.$
We first recall the relevant results from string theory.  
In particular we explain how the D-branes affect the integral
of K\"ahler form around 2-cycles surrounding D-branes and we
use this to formulate the D-branes in the crystal melting
problem as introducing defects.  We show that the amplitudes
for D-branes agree to all orders in string perturbation
with the melting crystal picture.  We also get additional non-perturbative
pieces, suggesting a non-perturbative completion of string
theory amplitudes.
\subsection{D-branes in topological A-model string theory on $\C$} 
One may view $\C$ as  a $T^3$ fibration.
  Let
$(z_1,z_2,z_3)$ denote complex coordinates
of $\C$ and let 
$$(x,y,z)=(|z_1|^2,|z_2|^2,|z_3|^2)$$ 
be the coordinates on
the base $ {\cal O}^{+}$ which is the positive octant in $R^3.$
We denote  the coordinates
on the fiber $T^3$ by $\tht_i, \ i=1,2,3.$
\begin{figure}
\begin{center}
\epsfig{file=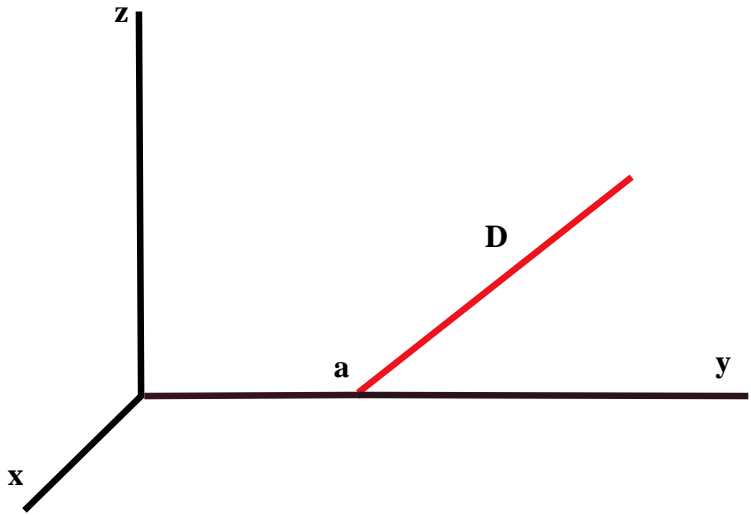,width=100mm}
\end{center}
\caption{ Projection to the base of a D-brane ending on y-axis at y=a. }
\label{db}
\end{figure}
Now consider a Lagrangian D-brane
which has the geometry given by
\be
y=x+a=z+a,\quad \quad \tht_1+\tht_2+\tht_3=0, \quad a>0.
\label{geom}
\ee
The D-brane has topology $S^1\times R^2$ and 
ends on the y-axis  at $y=a$ ( see Figure \ref{db}).
As discussed in \cite{AKMV} there is back-reaction of the geometry 
to the presence of the D-brane. Namely, the periods of the K\"ahler form
$K$ are changed by the amount
\be
\Delta \int_{\Sigma} K = g_s
\label{kahler}
\ee 
where $\Sigma$ is any 2-cycle linking the D-brane world-volume. 
This back-reaction is going to be our guide in search for crystal
interpretation of the brane.

\begin{figure}
\begin{center}
\epsfig{file=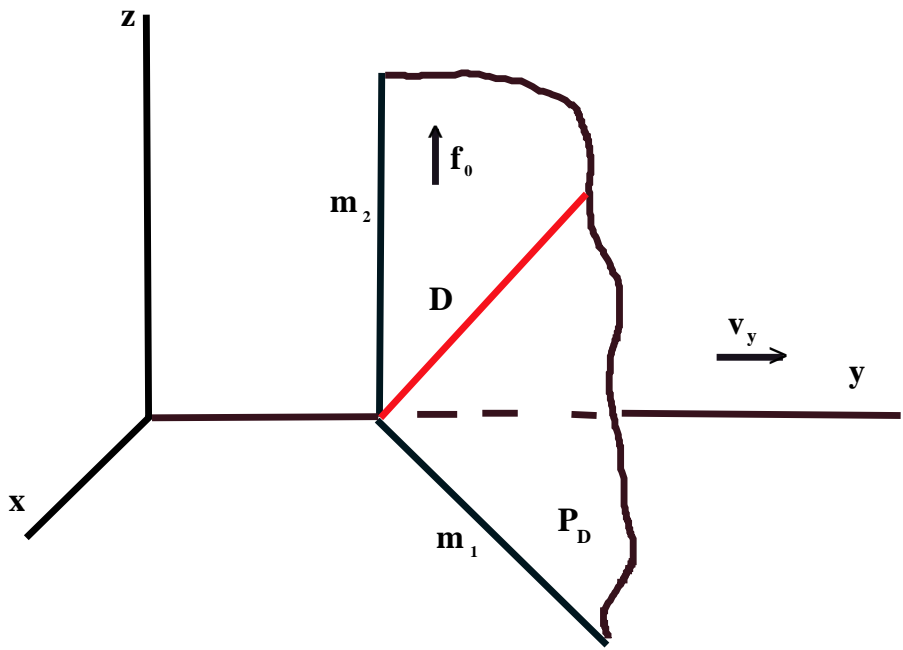,width=120mm}
\end{center}
\caption{  The D-brane with zero framing. Plane $P_D$ contains the projection $D$
of the D-brane to ${\cal O}^+.$ }
\label{fig9}
\end{figure}

Also, as explained in \cite{AKV}, 
in order to define quantum theory on a non-compact D-brane one has to 
specify boundary conditions at infinity called the ``framing''
of the D-brane. 
The trick is to modify the background by introducing
extra direction $\vec f$ in the toric base over which one of the cycles of the $T^3$ 
 fiber degenerates along a ``brane'' in that direction at infinity.   There is an integral
ambiguity in the choice of $\vec f$:  
$$\vec f\rightarrow \vec f -p \vec v_y$$
$p$ is called the framing number.
 This leads
to the notion of a preferred plane containing the D-brane.  Let's call this plane
the D-brane plane $P_D$.  A choice of the framing is also equivalent to the choice
of the plane $P_D$ containing the D-brane 
and the choice of one of the two directions $m_1,m_2$.  This follows
from the orientation of $\vec f$ versus $-\vec f$.
 The concept of framing is illustrated in Figure \ref{fig9}.   
We will show that the notion of framing enters  naturally 
in the crystal melting picture of the D-brane.

The choice of  $p=0$ framing in the convention of \cite{AKV}\footnote{ $p=-1$ in the
 convention of \cite{AKMV} } 
is the simplest one.
The partition function of the 
topological A-model string
theory on $\C$ in the presence of the D-brane (anti D-brane) 
with geometry (\ref{geom}) and
framing $p=0$
is given by \cite{OOV}
\be
Z_{D}^{string}(q,a)=M(q)\prod_{n=1}^{\infty} \Bigl( 1-e^{-a}q^{n-\half}\Bigr ),\quad \quad q=e^{-g_s}
\label{brane}
\ee
\be
Z_{\D}^{string}(q,a)=M(q)\prod_{n=1}^{\infty} \Bigl( 1-e^{-a}q^{n-\half}\Bigr )^{-1} 
\label{antibrane}
\ee
where $M(q)$ is the McMahon function (\ref{zclosed}).  Note that since
the brane amplitude at framing $p$ is equivalent to anti-brane amplitude
at framing $1-p$ the above equations can also be viewed as statement
for anti-brane (brane) with framing number 1.

Also, note that free energy $F=-logZ_{D}^{string}(q,a) $ has the following form:
\be
F=\sum_{n=1}^{\infty} \frac{e^{-na} }{ 2n {\rm sinh} (ng_s/2)}
\ee
so that perturbative expansion 
$
F_g\sim (g_s/a)^{2g-1} 
$
breaks down for $a \sim g_s.$
Nevertheless, the partition function $ Z_{D}^{string}(q,a)$ is well defined
for $a \sim g_s$ and it is natural to ask what this sum computes.
To find the answer to this question we are going to study the
interpretation of D-branes
in the melting crystal.
\subsection{D-branes as defects in the melting crystal}
How can one detect the presence of a D-brane in the crystal melting
picture? To answer this question we would like to use the important information
about the back-reaction of the geometry (\ref{kahler}).
Let us first recall  the role of the K\"ahler form $K$ in the crystal picture.
It was discovered in \cite{INOV} that
the melting crystal partition function $Z_{crystal}(q)$ arises  from
the sum over generalized\footnote{In a sense that the sum includes non-geometric excitations required
for a consistent quantum theory} K\"ahler geometries on $\C$ with K\"ahler forms $K$
quantized in units of $g_s.$ 
From this viewpoint the number of  atoms removed
from the corner can be thought of as  the change in the number 
of holomorphic sections of the $U(1)$ bundle whose curvature is $K/g_s.$
 
\begin{figure}
\begin{center}
\epsfig{file=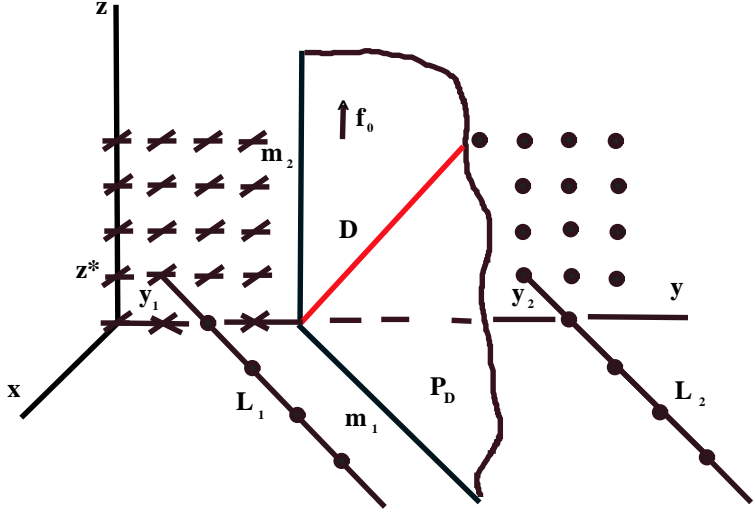,width=100mm}
\end{center}
\caption{
The half-lines $L_1$ and $L_2$  enter the definition of  the 2-cycle $\Sigma.$ 
 A cross  stands for vacancy and dot for atom.}
\label{fig2}
\end{figure}

Now  we introduce a D-brane of geometry (\ref{geom}).  The projection of the brane
to the base and the framing vector define a plane $P_D$  in ${\cal O}^+.$ 
This plane intersects $z=0$ plane along the half-line $m_1: \ y=a+(1+p)x$
and $x=0$ plane along the half-line $m_2: \ y=a-pz$ (see Figure \ref{fig9}).
There is a natural choice of a
 2-cycle linking this D-brane which we denote by $\Sigma$  
dictated by the choice of framing:  This cycle is a difference of planes
whose projection
 to the toric base gives a pair of half-lines $L_1,L_2$ parallel to $P_D$ in the
 direction of $m_1$ (fixed by the choice of framing).  We postulate that this gives rise to a `closed cycle' at infinity
 $\Sigma=L_2-L_1,$
  compatible with the fact that at infinity the framing of the
 D-brane chooses an extra
vanishing direction of a circle.
Here $L_1,L_2$ are given by
\be
L_1=(x,y_1+(1+p)x,z^*),\quad \quad L_2=(x,y_2+(1+p)x,z^*)
\label{lines}
\ee
where $y_1,y_2$ and $z^*$ are fixed nonnegative integers such that
$y_1 < a < y_2.$
The lines $L_1$ and $L_2$ together with $\theta_1$ angle give
rise to planes, which for general
$p$ are not holomorphically embedded in $\C$.
The period of the K\"aler form $K$ along the  2-cycle $\Sigma$
is  given by
\be 
\int_{\Sigma}K=g_s (\vert L_2\vert -\vert L_1\vert),
\label{intlinesii}
\ee
where $\vert L_i\vert $ is the number of atoms on the half-line $L_i,\quad i=1,2.$
From (\ref{intlinesii}) follows that the change of the period by one in $g_s$ units, 
which is the characteristic signature of the D-brane, can be achieved by, for example, 
dropping a single atom from the half-line  $L_1$.  
We illustrate this idea in Figure \ref{fig2}.

\begin{figure}
\begin{center}
\epsfig{file=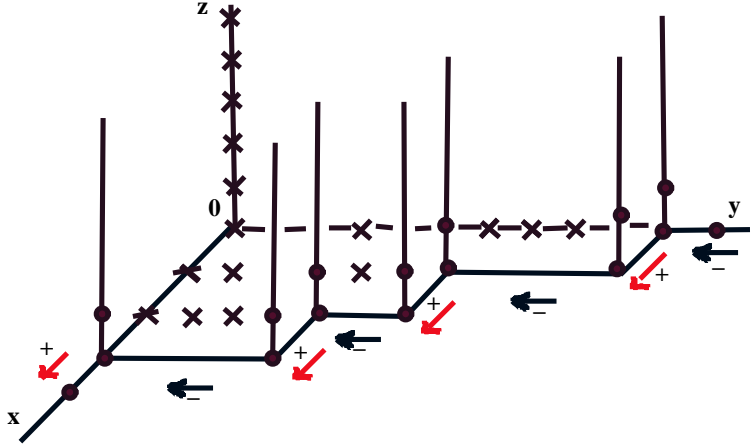,width=100mm}
\end{center}
\caption{The shape of the melting crystal in the presence of D-branes.
 The  arrows carry negative
(positive) sign and show 
the evolution with $\Gamma_{-}$ ($\Gamma_{+}$). The crosses  are vacant sites inside the
cylinder ${\cal M}_{\nu}.$ 
}
\label{alt}
\end{figure}

Recalling that parameters $y_1,y_2,z^*$ which specify a linking 2-cycle
$\Sigma$ are arbitrary nonnegative
integers ( with the only constraint $ y_1 < a < y_2$ ) we see
that in the presence of a D-brane one layer
of atoms is deleted from the $x=0$ plane, in the region $0\leq y < a$.
In other words the crystal develops an extra 
corner. Thus,  we are led
to the interpretation of D-branes as defects modifying 
the shape of the crystal.  For many D-branes the crystal
will develop many corners.   Moreover the concavity or convexity
of the corner is correlated with whether we have a D-brane or anti-D-brane.
See Figure \ref{alt}.

Note that the picture of modified crystal with D-branes at various framings
makes sense as long as the corresponding $m_2$ lines do not intersect the crystal
axis $z$ or the other corners developed.  Thus only for these cases
we can propose a simple dual crystal interpretation.  Moreover the notion
of linking cycle $\Sigma$ that we defined will only make sense for two choices
of framings for each brane ($p=0,-1$) otherwise the corresponding cycles intersect
the boundary of the crystal.  This already suggests that perhaps
the rule of crystal melting may be different for general framings
and this indeed is consistent with what we shall find.  In fact the simplest
rule is for framing $p=0$ and that is the case we focus on.

One would naturally propose 
that the modification due to the D-brane in the A-model partition function
is the same as the statitical mechanical model
of melting crystal but now with the modified crystal
 $O^+/{\cal M}_{\nu}$ studied in \cite{OR,ORV}.
Here  ${\cal M}_{\nu}$ is a cylinder with the 2d partition $\nu$ as its base.
In other words, one  considers 3d partitions in  ${\cal O}^+$ and 
regards the two 3d partitions as identical if they differ only
inside  ${\cal M}_{\nu}.$ The weight of the 3d partition is 
determined by the number of boxes outside ${\cal M}_{\nu}.$  We will see
that this is indeed the correct rule for the case of  framing $p=0$.  For
general framing the crystal geometry is modified as we have found, but
the melting rules also need to be modified.  Some aspects of this will be
discussed in the next section.  In the remainder of this section we concentrate
on the case with $p=0$ and show how to recover the D-brane amplitudes
from the melting crystal.

In the operator language the 3d partition
in the modified container is obtained from alternating evolution
with $\Gamma_{-}$ and $\Gamma_{+}$ whose action on diagonal slices provide appropriate interlacing
pattern. We illustrate the evolution giving ``room with the corners ''
 in Figure \ref{alt}.      
Now we would like to check
 the proposed interpretation for the D-branes  by explicit computation
using realization of $\Gamma_{\pm}$ as operators in fermion Hilbert space.  It turns out that this realization of D-branes
is consistent with the
intuition developed in the B-model topological string theory \cite{AKMV} that non-compact D-branes
are fermions.

Let us recall \cite{OR} the representation of operators $\Gamma_{\pm}$ in the
bosonized picture.
One introduces chiral boson $\phi(z)$ such that
\be
\psi^*(z)=: e^{-\phi}(z):, \quad   \psi(z)=: e^{\phi}(z):
\label{bozon}
\ee
and expand $\phi(z)$ into zero, positive and negative modes
\be
\phi(z)=\phi_0(z)+\phi_{+}(z)+ \phi_{-}(z),\quad
\bigl [\phi_{+}(z),\phi_{-}(w) \bigr ]=log(1-\frac{z}{w})
\label{expand}
\ee
Then $\Gamma_{\pm}(z)$  are given by
\be
\Gamma_{\pm}(z)=exp\Biggl( \pm \phi_{\pm}(z) \Biggr )
\label{defgam}
\ee
Note that $\psi(z)$ and $\psi^*(z)$ are related to $\Gamma_{\pm}(z)$ :
\be
\psi^*(z)=e^{-\phi_0} \Gamma_{-}(z)\Gamma_{+}^{-1}(z), \quad
\psi(z)=e^{\phi_0} \Gamma_{-}^{-1}(z)\Gamma_{+}(z)
\label{relation}
\ee
Now we propose that introducing the D-brane of the geometry (\ref{geom})
at $a=g_s(N_0+\half)$ with framing p=0 amounts to 
inserting the operator 
\be
\Psi_D(z)=\Gamma_{-}^{-1}(z)\Gamma_{+}(z)
\label{psid}
\ee  
at $z=q^{-(N_0+\half)}$ in the correlator  (\ref{operii}).
We choose not to include the zero mode part $e^{\phi_0} $ in the definition
of the D-brane operator $\Psi_D(z)$ since from the crystal  viewpoint it is more natural 
to deal with zero fermion number correlators. 
As discussed in \cite{AKMVb} disregarding zero modes corresponds to having compensating
(fermion number ) flux at infinity. 

To verify our proposal for the D-brane operator let us compute
 \be
Z_{D}^{crystal}(q,N_0):= <0 \vert \prod_{n=1}^{\infty}\Gamma_{+}\bigl (q^{n-\half}\bigr ) 
\prod_{m=1}^{N_0+1}\Gamma_{-}\bigl ( q^{-m+\half}\bigr )  
\Psi_D\bigl (q^{-(N_0+\half)}\bigr )
\prod_{N_0+2}^{\infty}\Gamma_{-}\bigl ( q^{-m+\half}\bigr )\vert 0>
\label{doperii}
\ee
 Using our definition (\ref{psid}) we have 
\be
Z_{D}^{crystal}(q,N_0)=<0 \vert \prod_{n=1}^{\infty}\Gamma_{+}\bigl (q^{n-\half}\bigr ) 
\prod_{m=1}^{N_0}\Gamma_{-}\bigl ( q^{-m+\half}\bigr )  
\Gamma_{+}\bigl (q^{-(N_0+\half)}\bigr )
\prod_{N_0+2}^{\infty}\Gamma_{-}\bigl ( q^{-m+\half}\bigr )\vert 0>
\label{newoperii}
\ee
Note that insertion of a D-brane $\Psi_D$ creates an anticipated disorder at $q^{-(N_0+\half)}$ in a sequence
of $\Gamma_{-}$ operators.
The commutation relation (\ref{gamcom}) gives the result
\be
 Z_{D}^{crystal}(q,N_0)=M(q)\xi(q)\prod_{n=1}^{\infty} \Bigl( 1-e^{-g_s (N_0+\half)} q^{n-\half}\Bigr ),
\quad q=e^{-g_s}
\label{newres}
\ee
where $\xi(q)=\prod_{n=1}^{\infty} (1-q^n)^{-1}.$
Let us compare string answer (\ref{brane}) with (\ref{newres}):
\be
Z_{D}^{crystal}(q,N_0)= \xi(q)Z_D^{string}(q,a),\quad a=g_s(N_0+\half)
\label{compare}
\ee
The renormalization factor $\xi(q)$ is important for the melting crystal
picture of the D-brane.
Due to $\xi(q)$ factor we can write  $Z_{D}^{crystal}(q,N_0)$ in the simple form:
\be
 Z_{D}^{crystal}(q,N_0)=M(q)\prod_{n=1}^{N_0} ( 1- q^{n} )^{-1}=P\bigl(0,0,\nu(N_0)\bigr)(q)
\label{newresii}
\ee
Here $\nu(N_0)$ stands for the 2d partition with single row of length $N_0$ and
 $P\bigl(0,0,\nu \bigr)(q)$  was defined in \cite{ORV} as   
  the partition function of the 3d partitions in the
modified container ${\cal O}^+/{\cal M}_{\nu}.$

\begin{figure}
\begin{center}
\epsfig{file=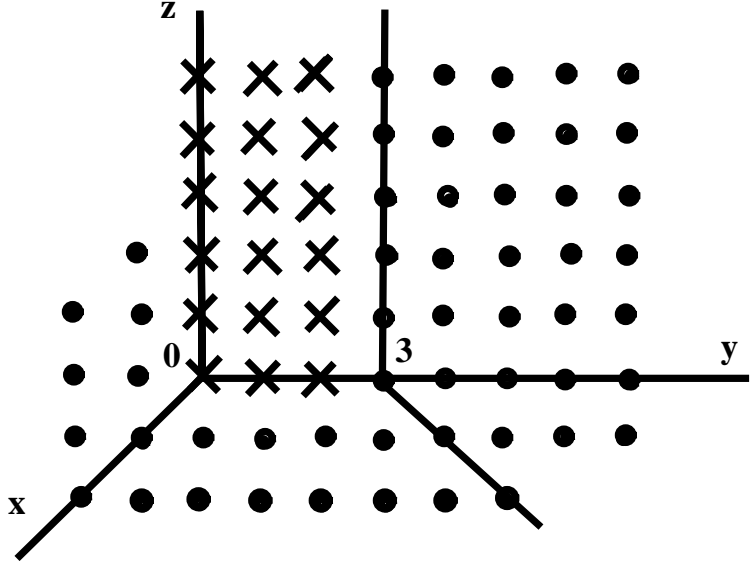,width=100mm}
\end{center}
\caption{ Modified lattice for a D-brane with $p=0, N_0=3.$ The crosses stand for vacancies and dots for atoms.}
\label{fig4}
\end{figure}
$P\bigl (0,0,\nu(N_0) \bigr)(q)$ can also be viewed as the partition function
of the melting crystal when melting starts from the cubic lattice  $\Lambda \in  {\cal O}^+$
with all the sites  
$$(0,y,z), \quad y < N_0 g_s$$
 already vacant  (see Figure \ref{fig4} ).

We now easily generalize the above result to give  melting crystal picture  for the M D-branes of the geometry (\ref{geom})
with $a_i=g_s(N_i+\half),\quad i=1,\ldots,M$ and $ N_M < N_{M-1}<\ldots N_1.$
In the operator language we compute
\be
Z_{M D-branes}^{crystal}(q,N_i):= <0\vert \prod_{n=1}^{\infty}\Gamma_{+}\bigl (q^{n-\half}\bigr ) 
\prod_{m=1}^{N_M +1}\Gamma_{-}\bigl ( q^{-m+\half}\bigr )  
\Psi_D\bigl (q^{-(N_M+\half)}\bigr )
\label{opertwod}
\ee
$$ \prod_{m=N_M+2}^{N_{M-1}+1}\Gamma_{-}\bigl ( q^{-m+\half}\bigr ) \Psi_D\bigl (q^{-(N_{M-1}+\half)}\bigr )
\prod_{m=N_{M-1}+2}^{N_{M-2}+1}\Gamma_{-}\bigl ( q^{-m+\half}\bigr )\ldots
$$
$$\ldots 
\prod_{m=N_2+2}^{N_{1}+1}\Gamma_{-}\bigl ( q^{-m+\half}\bigr ) \Psi_D\bigl (q^{-(N_1+\half)}\bigr )
\prod_{m=N_{1}+2}^{\infty}\Gamma_{-}\bigl ( q^{-m+\half}\bigr )\vert 0>
$$

\begin{figure}
\begin{center}
\epsfig{file=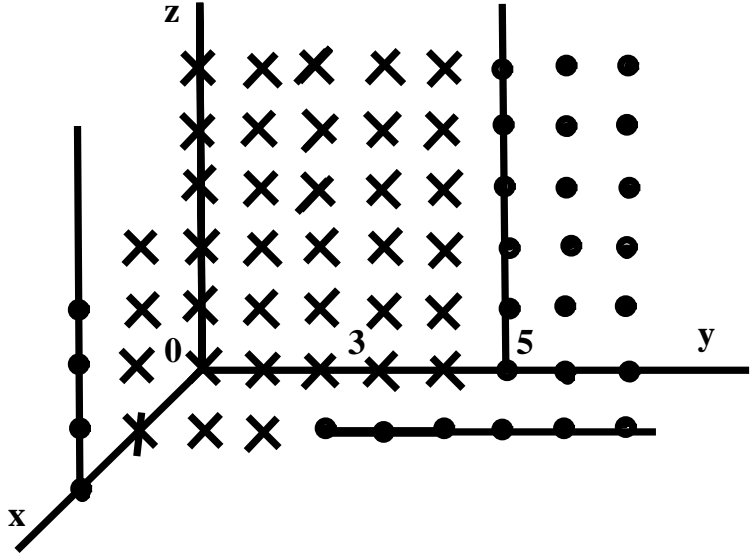,width=100mm}
\end{center}
\caption{ Modified lattice for two D-brane with $p=0$ and $N_1=6,N_2=3.$}
\label{fig5}
\end{figure}

Evaluating the above correlator using (\ref{gamcom} ) we find
\be
Z_{M D-branes}^{crystal}(q,N_i)=
M(q)\prod_{i<j} \Bigl(1-q^{N_i-N_j}\Bigr )\prod_{i=1}^M\prod_{n_i=1}^{N_i} ( 1- q^{n_i} )^{-1}
=
P\Bigl( 0,0, \nu\bigl( N_i \bigr ) \Bigr)
\label{simple}
\ee
where $\nu\bigl(N_i\bigr)$ stands for the 2d partition such that $\nu_i=N_i-M+i.$
 The last equality in (\ref{simple}) follows from the
relation established in \cite{ORV}
\be
P\bigl( 0,0, \nu \bigr)(q)=M(q)q^{-\frac{\vert \vert \nu^T \vert \vert^2}{2}}C_{..\nu}(q^{-1})
\ee
where $\vert \vert \lam \vert \vert^2 = \sum_i \lam_i^2$ and $C_{..\nu}(q^{-1})$
is the topological vertex defined in \cite{AMV}.
In the crystal language one is melting the cubic lattice  in the positive octant
with all the sites 
$$\bigl(g_s(i-1),y,z \bigr), \quad y < g_s\bigl( N_i-M+i \bigr),\quad i=1,\ldots,M$$ 
already vacant (see Figure \ref{fig5} for the case of two D branes).

We have considered D-branes  which end on the y-axis. 
Let us now insert an anti-D-brane ending at $x=g_s(N_0+\half).$
We claim that the anti-D-brane operator is  given by 
\be
\Psi_{{\overline D},x}(z)=\Gamma_{-}(z)\Gamma_{+}^{-1}(z)
\label{psidx}
\ee
inserted at 
$z=q^{(N_0+\half)}$
where  subscript $x$  indicates that we are dealing with anti-D-branes
ending at x-axis.  This is clear from the jumping rule for the K\"ahler
class.  We can now check this claim by computing the amplitude:
\be
Z_{{\overline D},x}^{crystal}(q,N_0):= <0 \vert  
\prod_{m=N_0+2}^{\infty}\Gamma_{+}\bigl ( q^{m-\half}\bigr )  
\Psi_{{\overline D},x}\bigl (q^{(N_0+\half)}\bigr )\prod_{m=1}^{N_0+1}\Gamma_{+}\bigl ( q^{m-\half}\bigr )  
\prod_{n=1}^{\infty}\Gamma_{-}\bigl ( q^{-n+\half}\bigr )\vert 0>
\label{yoperiii}
\ee
We  use (\ref{psidx}) to recast $Z_{{\overline D},x}^{crystal}(q,N_0)$ as
\be
Z_{{\overline D},x}^{crystal}(q,N_0)= <0 \vert  
\prod_{m=N_0+2}^{\infty}\Gamma_{+}\bigl ( q^{m-\half}\bigr )  
\Gamma_{-}\bigl (q^{(N_0+\half)}\bigr )\prod_{m=1}^{N_0}\Gamma_{+}\bigl ( q^{m-\half}\bigr )  
\prod_{n=1}^{\infty}\Gamma_{-}\bigl ( q^{-n+\half}\bigr )\vert 0>
\label{yoperiiiv}
\ee
Note that insertion of  $\Psi_{{\overline D},x}$ creates 
a disorder at $q^{(N_0+\half)}$ in a sequence
of $\Gamma_{+}$ operators.
Computing (\ref{yoperiiiv}) we find 
\be
Z_{{\overline D},x}^{crystal}(q,N_0)=M(q)\xi(q)\prod_{n=1}^{\infty} 
\Bigl( 1-e^{-g_s (N_0+\half)} q^{n-\half}\Bigr ),
\quad q=e^{-g_s}
\label{ynewres}
\ee
where $\xi(q)=\prod_{n=1}^{\infty} (1-q^n)^{-1}.$
Note that (\ref{ynewres}) is equal ( up to the factor $\xi(q)$ )
 to the string answer (\ref{brane}) for
anti-D-brane with framing $p=1.$

\begin{figure}
\begin{center}
\epsfig{file=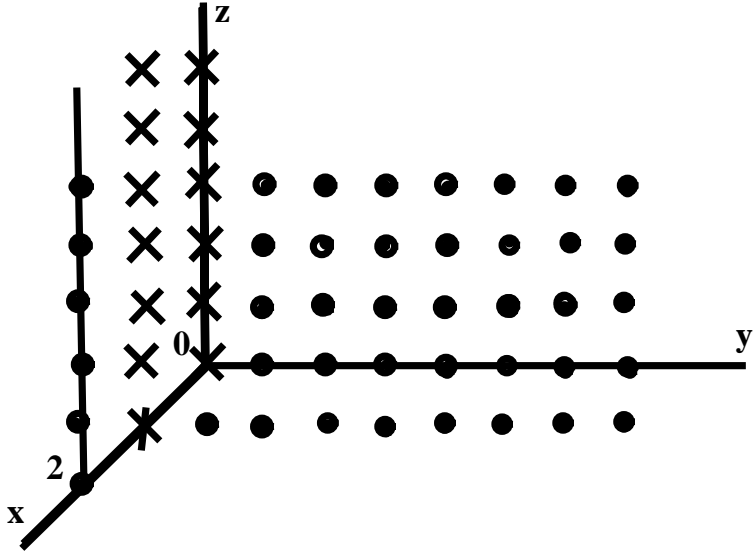,width=100mm}
\end{center}
\caption{ Modified lattice ${\overline \Lambda}_x$ for anti-D-brane with $p=1, N_0=2$ ending at x-axis.}
\label{fig6}
\end{figure}

$Z_{{\overline D},x}^{crystal}(q,N_0)$ can also be viewed as the partition function
of the melting crystal when melting starts from the  lattice 
${\overline \Lambda}_x \in {\cal O}^+$  
with all the sites 
$$(x,0,z), \quad x < N_0 g_s$$
 already vacant (see Figure \ref{fig6}). 

We would like to end this section with a comment about extra factor $\xi(q)$
which appears in the crystal melting picture as compared with
the string theory answer (\ref{compare}). To ensure that this extra factor cannot be detected
in perturbative stringy computations we propose to
modify the definition of a D-brane operator in crystal picture
\be
{\widehat{ \Psi_D}}=q^{-1/24} \Psi_D
\label{newdef}
\ee
Then, extra factor would be $\eta(q)^{-1}$ and the change in the free
energy due to this factor is non-perturbative in the $g_s \to 0$ limit.
\be
\Delta F=-log \eta(q)=- log \eta({\tilde q})+\half log \frac{g_s}{2\pi},\quad 
\quad {\tilde q}=exp\bigl ( -\frac{(2\pi)^2}{g_s} \bigr)
\label{nonpert}
\ee 
In the S-dual language \cite{NOV} the factor  $\eta(q)^{-1}$ may be thought as coming
from extra $D(-1)$ instantons in the presence of Lagrangian NS2-brane.   It would be
interesting to determine analogs of $\xi(q)$ for arbitrary framings.
 
\section{Crystal amplitudes for more general D-brane configurations}

So far we have discussed crystal melting interpretation for special
configuration of D-branes.  Here we would like to consider more
general configuration of branes and anti-branes at various framings.

\subsection{Anti-D-brane amplitudes}

\begin{figure}
\begin{center}
\epsfig{file=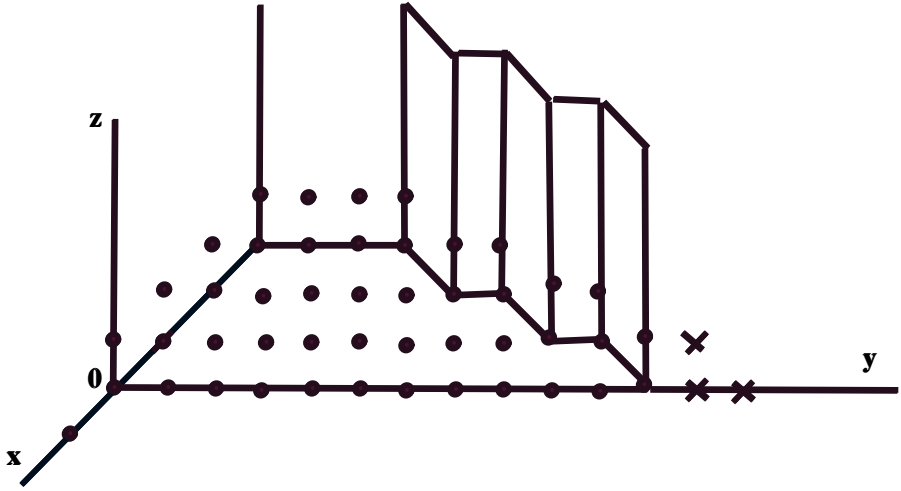,width=120mm}
\end{center}
\caption{ The shape of the melting crystal in presence of anti-D-branes.}
\label{anti}
\end{figure}

Let us consider
anti-D-brane ending on the $y$ axis and zero framing.
We will take into consideration that the jump of the K\"ahler period along
the 2-cycle linking anti-D-brane is opposite in sign to the corresponding
jump around D-brane.
Then, we can naturally propose the shape of the melting crystal
as shown in Figure \ref{anti}.   Note that the corners are now concave,
as opposed to the case for branes ending on the $y$-axis.

\begin{figure}
\begin{center}
\epsfig{file=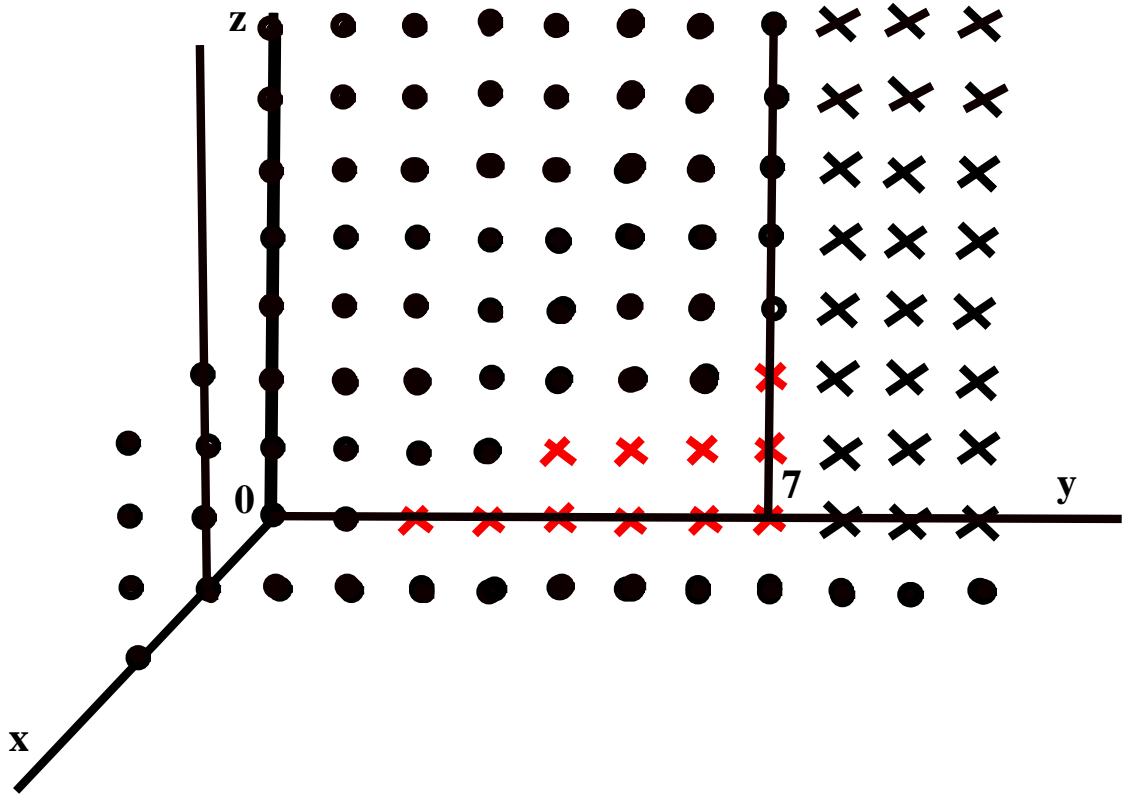,width=150mm}
\end{center}
\caption{ An example of an allowed configuration molten in $x=0$ plane.
The number of columns is three, $N_0=8,$ the contribution to $Z$ is $(-1)^3 q^{11}.$
 }
\label{fig7}
\end{figure}

To verify this proposal and find the rules of melting we
again use operator language. 
We suggest that  anti-D-brane ending
on y-axis at $y=g_s(N_0+\half)$ is described by insertion
of the operator 
\be
\Psi_{\D}(z)=\Gamma_{-}(z)\Gamma_{+}^{-1}(z)
\label{psiantid}
\ee
at $z=q^{-(N_0+\half)}.$

\be
Z_{\D}^{crystal}(q,N_0):= <0 \vert \prod_{n=1}^{\infty}\Gamma_{+}\bigl (q^{n-\half}\bigr ) 
\prod_{m=1}^{N_0+1}\Gamma_{-}\bigl ( q^{-m+\half}\bigr )  
\Psi_{\D}\bigl (q^{-(N_0+\half)}\bigr )
\prod_{N_0+2}^{\infty}\Gamma_{-}\bigl ( q^{-m+\half}\bigr )\vert 0>
\label{operiii}
\ee
We  use (\ref{psiantid}) to recast $Z_{\D}^{crystal}(q,N_0)$ as
\be
 <0 \vert \prod_{n=1}^{\infty}\Gamma_{+}\bigl (q^{n-\half}\bigr ) 
\prod_{m=1}^{N_0}\Gamma_{-}\bigl ( q^{-m+\half}\bigr ) \Gamma_{-}^2\bigl (q^{-(N_0+\half)}\bigr ) 
\Gamma_{+}^{-1}\bigl (q^{-(N_0+\half)}\bigr )
\prod_{N_0+2}^{\infty}\Gamma_{-}\bigl ( q^{-m+\half}\bigr )\vert 0>
\label{operiianti}
\ee
Note that insertion of an anti-D-brane $\Psi_{\D}$ creates a much more complicated disorder at $q^{-(N_0+\half)}$ in a sequence
of $\Gamma_{-}$ operators than D-brane does.
The commutation relation (\ref{gamcom}) gives the result for anti-D-brane
\be
 Z_{\D}^{crystal}(q,N_0)=M(q)\xi^{-1}(q)\prod_{n=1}^{\infty}
 \Bigl( 1-e^{-g_s (N_0+\half)} q^{n-\half}\Bigr )^{-1}=M(q)\prod_{n=1}^{N_0} ( 1- q^{n} ),
\quad q=e^{-g_s}
\label{newresanti}
\ee
Let us compare string answer (\ref{antibrane}) with (\ref{newresanti}):
\be
Z_{\D}^{crystal}(q,N_0)= \xi^{-1}(q)Z_{\D}^{string}(q,a),\quad a=g_s(N_0+\half)
\label{compareanti}
\ee
Now we would like to give a crystal  interpretation of (\ref{newresanti}).
A natural guess which gives the correct jump
of the K\"ahler class is to start melting the cubic lattice
$\overline \Lambda $ in the positive octant with vacant sites 
$$(0,y,z), \quad y \ge g_s N_0 .$$  It turns out that in addition
to the choice of this modified lattice,
 one also has to modify the melting rule
to reproduce the correct answer predicted from the operator formulation
(and known string amplitudes).
To formulate the modified rule we use  
\be
\prod_{n=1}^{N_0} (1-q^n)=\sum_{\nu \in {\cal P}_{N_0} }
(-)^{ \sum_{i=1}^{N_0} n_i} q^{\sum_{i=1}^{N_0} i n_i}
\label{columnstrict}
\ee
where ${\cal P}_{N_0}$ is a set of column
 strict\footnote{ Column strict means that the 
 lengths of columns  are strictly decreasing }
2d partitions with maximum column length $ l_{max} \le N_0$ and $n_i=0,1$
 counts if a column of length $i$ is present in $\nu$.
Here is a tentative proposal for the modified rule.
The melting in $x=0$ plane starts from $\bigl(0,(N_0-1)g_s,0\bigr)$ and is going 
in the direction of decreasing y and increasing z.
The allowed configurations for this new melting
are collections of atoms forming column strict
2d partitions. So for example, a configuration
$$ \Bigl \{ \bigl(0,(N_0-1)g_s ,0\bigr ), \bigl (0,(N_0-1)g_s,g_s \bigr) \Bigr \}$$
is not allowed
 but a
configuration
\be
\Bigl \{ \bigl (0,(N_0-1)g_s,0 \bigr), \bigl(0,(N_0-2)g_s,0\bigr),\bigl(0,(N_0-1)g_s,g_s\bigr)
\Bigr \} 
\label{config}
\ee 
is allowed.
The sign of configuration is the number of columns,
which is two in the example (\ref{config}).
We give a more nontrivial example of an allowed configuration
molten in the $x=0$ plane in Figure \ref{fig7}.
The melting in the rest of the positive octant goes in a standard way (\ref{rules})
independently of the melting in the $x=0$ plane.
For this reason a total configuration of  molten atoms  
has the form $(A; B)$
where A is a configuration obtained from melting in ${\cal O}^+$
and B is a configuration (with signs) 
 in the $x=0$ plane. 
For more anti-D-branes or mixtures of D-branes
and anti-D-branes, one naturally expects that the modified
lattice be consistent with the change in K\"ahler class.  Viewing
D-brane as fermion and anti-D-brane as anti-fermion, this is also
consistent with the fact that $\oint \partial \phi$ measures the K\"ahler class
of the crystal and its jump across the insertion of fermion or anti-fermion
is consistent with the expected geometry of the lattice.  However,
the precise rule for the melting crystal remains to be worked
out for this more general melting.  Of course the operator expressions
always exist for such amplitudes. In this regard, we would like to emphasize that the global
rule of melting is also absent for closed strings on toric Calabi-Yau manifolds
with more than one fixed point of the toric action. The rules given in \cite{INOV}
are local\footnote {There are also modifications which account for world sheet instantons
wrapped on compact 2-cycles between fixed points} near each of the fixed point.  This is related to the fact that what appears
as brane to one corner of the crystal will appear as anti-brane to
another corner.
In the spirit of \cite{INOV}  we can  get a rough  idea about the melting crystal
interpretation  of the
D-brane and anti-D-brane  when they are far apart.
In this case we can regard our 
models of melting as local ones which are valid in the vicinity of each brane.

\begin{figure}
\begin{center}
\epsfig{file=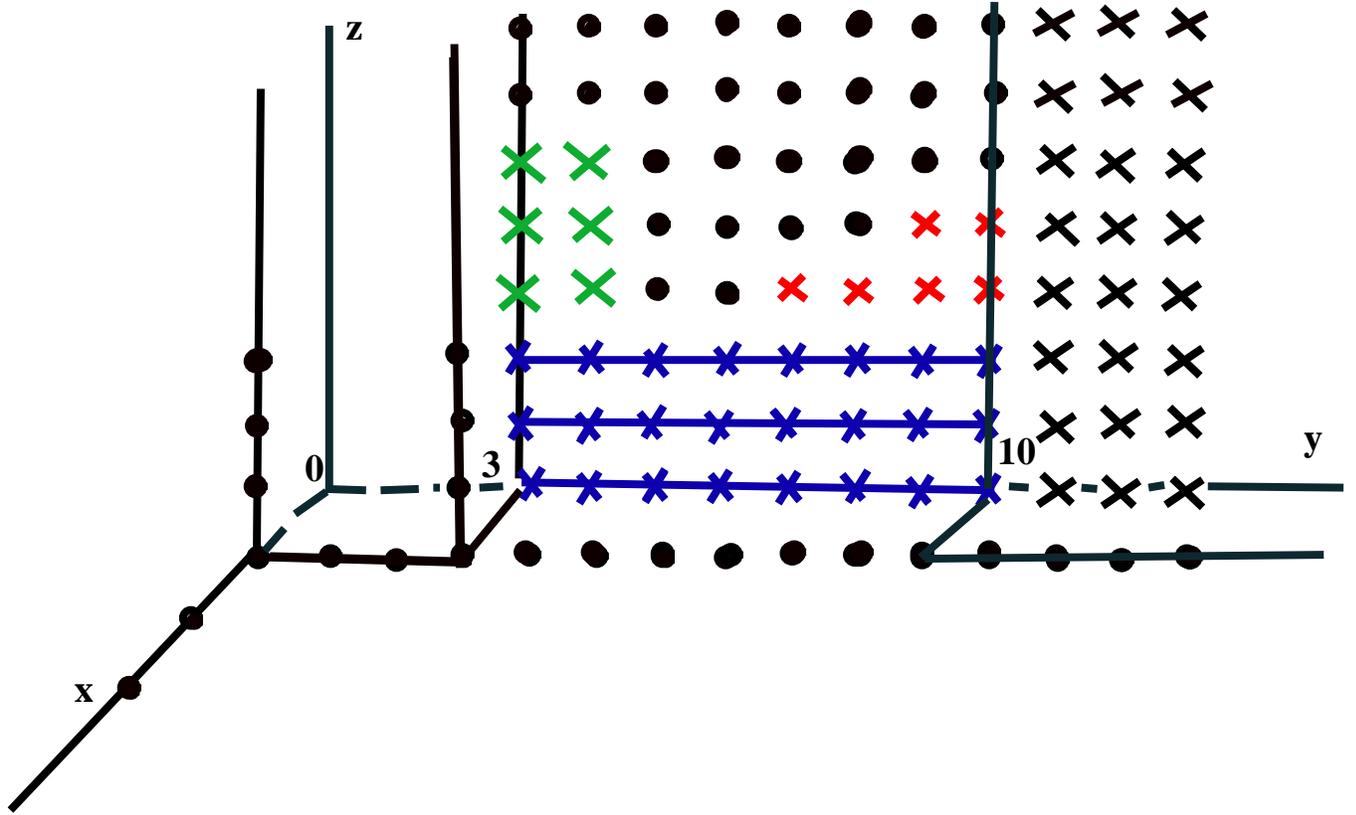,width=180mm}
\end{center}
\caption{ $N_1=3,\quad N_2=11.$ \ 6 atoms are molten  near each of the branes on top of the three molten strings. }
\label{fig8}
\end{figure}
Let us consider  a D-brane at $a=g_s(N_1+\half)$ and an anti-D-brane at $b=g_s(N_2+\frac{1}{2})$
so that $N_2 \gg N_1.$
Note that in the presence of D-brane and anti-D brane there are
essentially new types of objects (one dimensional lines) to be added
to the melting picture \cite{INOV}. These are ``strings of atoms `` of length
\footnote{the number of atoms in a string}
$N_2-N_1.$ The  new configurations account for the factor $(1-q^{N_2-N_1})^{-1}$
present in the partition function:
\be
Z_{D \ \D}=M(q)(1-q^{N_2-N_1})^{-1}\prod_{n=1}^{N_1}
(1-q^n)^{-1}\prod_{n=1}^{N_2}(1-q^m)
\label{pair}
\ee
We show in the Figure \ref{fig8} a typical configuration of molten atoms
on top of melting three  string-like objects. 
 
\subsection{The D-brane framing in crystal picture}
Here we would like to discuss the crystal amplitudes for
D-branes in arbitrary framing.  We will show, very much
like the case for the anti-D-branes, that the operator
formulation of the amplitude is compatible with the modified
lattice we had obtained before.  Though a direct interpretation
of the statistical mechanical model encoded by the operator
formulation is unclear.

Let us recall the string theory 
 partition function in the presence of  a D-brane (anti-D-brane)
  of the geometry (\ref{geom})
with framing $p$ \cite{AKMV}\footnote{We use convention for 
framing as in \cite{AKV}}.
\be
Z_{D}^{string}(q,a,p)=M(q)\sum_{k=0}^{\infty}e^{-ka} (-1)^{(1-p)k}q^{(1-p) 
k(k-1)/2}
q^{k/2}\prod_{n=1}^{k}(1-q^n)^{-1},\quad \quad q=e^{-g_s}
\label{pbrane}
\ee
\be
Z_{\D}^{string}(q,a,p)=M(q)\sum_{k=0}^{\infty}e^{-ka}(-1)^{pk} q^{p k(k-1)/2}
q^{k/2}\prod_{n=1}^{k}(1-q^n)^{-1},
\label{pantibrane}
\ee

\begin{figure}
\begin{center}
\epsfig{file=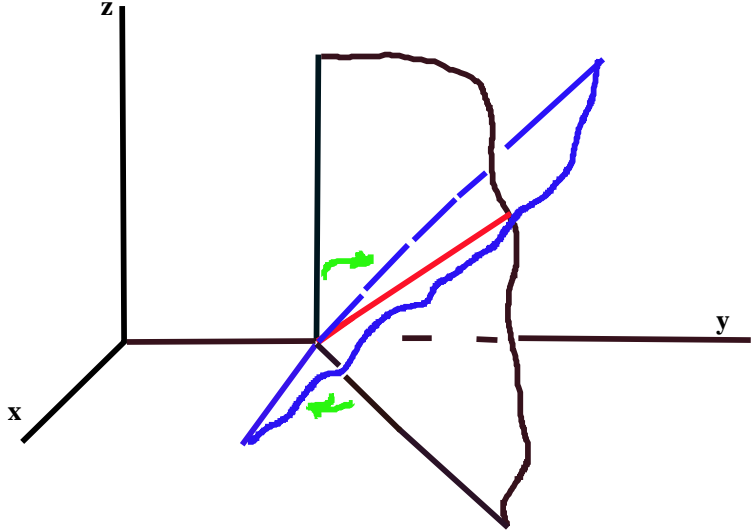,width=100mm}
\end{center}
\caption{$U_{-p}$ rotates 2d partition around its diagonal.}
\label{fig10}
\end{figure}

\begin{figure}
\begin{center}
\epsfig{file=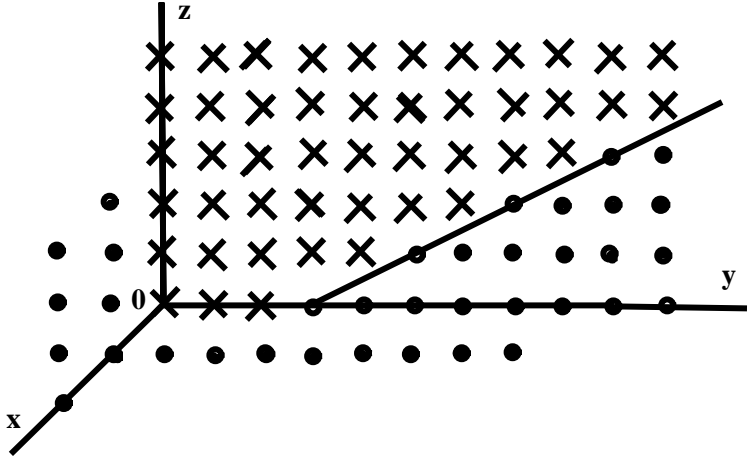,width=100mm}
\end{center}
\caption{Initial  lattice $\Lambda_{(-2)} $ for a D-brane with $p=-2, N_0=3.$}
\label{fig11}
\end{figure}

Now we are going to illustrate how string answers can be reproduced from
the operator formulation on the modified
crystal  (up to the renormalization factors $\xi^{\pm 1}$).
We define the fermion operators $\psi^{(p)}$ and $\psi^{*(p)}$ :
\be
\psi^{(p)}(z)=\sum_{n\in {\bf Z}} \psi_{n+\half}(-)^{p(n+1)}q^{-pn(n+1)/2} z^{-n-1},
\
\psi^{*(p)}(z)=-\sum_{n\in {\bf Z}} \psi^{*}_{n+\half}(-)^{p(n+1)}q^{pn(n+1)/2} z^{-n-1}
\label{ppsi}
\ee
and in analogy with (\ref{doperii},\ref{operiii})\footnote{ It is more convenient
here to insert  operators with nonzero fermion number  and thus change the vacuum at $t=\infty.$
For a single D-brane (anti-D-brane) this is equivalent to our treatment 
in section 3.} insert them at $q^{-(N_0+\half)}$
\be
Z_{D}^{crystal}(q,N_0,p):= <\bv_1 \vert \prod_{n=1}^{\infty}\Gamma_{+}\bigl (q^{n-\half}\bigr ) 
\prod_{m=1}^{N_0+1}\Gamma_{-}\bigl ( q^{-m+\half}\bigr )  
\psi^{(p)}\bigl (q^{-(N_0+\half)}\bigr )
\prod_{N_0+2}^{\infty}\Gamma_{-}\bigl ( q^{-m+\half}\bigr )\vert 0>
\label{poperii}
\ee
\be
Z_{anti-D}^{crystal}(q,N_0,p):= <\bv_{-1} \vert \prod_{n=1}^{\infty}\Gamma_{+}\bigl (q^{n-\half}\bigr ) 
\prod_{m=1}^{N_0+1}\Gamma_{-}\bigl ( q^{-m+\half}\bigr )  
\psi^{*(p)}\bigl (q^{-(N_0+\half)}\bigr )
\prod_{N_0+2}^{\infty}\Gamma_{-}\bigl ( q^{-m+\half}\bigr )\vert 0>
\label{poperiii}
\ee
where $<0\vert =<\bv_1 \vert e^{\phi_0}$ and $<0\vert =<\bv_{-1} \vert e^{-\phi_0}.$ 
In order to compute (\ref{poperii}) we first find correlators involving modes $\psi_{n+\half}$:
\be
{\cal P}_n=<\bv_1 \vert \prod_{n=1}^{\infty}\Gamma_{+}\bigl (q^{n-\half}\bigr ) 
\prod_{m=1}^{N_0+1}\Gamma_{-}\bigl ( q^{-m+\half}\bigr )  
\psi_{n+\half}
\prod_{N_0+2}^{\infty}\Gamma_{-}\bigl ( q^{-m+\half}\bigr )\vert 0>
\label{ppoperii}
\ee
These can be extracted from the known expression for $Z_{D}^{crystal}(q,N_0,p=0)$
(\ref{newresii}) to be
\be
{\cal P}_{n=-1+k}=M(q)\xi(q) (-)^k q^{k^2/2} \prod_{s=1}^{k}(1-q^s)^{-1}, \quad k\ge 0, \quad
{\cal P}_{n}=0,\quad n\le -2
\label{pin}
\ee
For anti-D-brane the procedure is analogous.
Then we immediately find the relation between crystal and string partition 
functions
\be
Z_{D}^{crystal}(q,N_0,p)= \xi(q)Z_D^{string}(q,a,p),\quad a=g_s(N_0+\half)
\label{pcompare}
\ee
\be
Z_{\D}^{crystal}(q,N_0,p)= \xi^{-1}(q)Z_{\D}^{string}(q,a,p),\quad a=g_s(N_0+\half)
\label{pcompareanti}
\ee
where $\xi(q)=\prod_{n=1}^{\infty} (1-q^n)^{-1}.$
Now an important point is that we can write operators $\psi^{(p)}$ and 
$\psi^{*(p)}$ as \cite{AKMV, AKMVb}
\be
\psi^{(p)}=U_p \psi U_{-p},\quad \psi^{*(p)}=U_p \psi^{*} U_{-p}
\label{conj}
\ee
Here $U_p=(-1)^{pL_0} q^{p C_2}$ and
\be
L_0=\sum_{k \in {\bf Z}} (k+1) :\psi_{k+\half}\psi^*_{-k-\half}:,
\quad C_2=-\sum_{k \in {\bf Z}} \frac{k(k+1)}{2} :\psi_{k+\half}\psi^*_{-k-\half}:
\label{import}
\ee
 Recall that $L_0$ and $C_2$ act  on the state  $\vert \mu >$ corresponding to 2d partition $\mu$ 
as follows:
\be
L_0 \vert \mu > = \vert \mu \vert \vert \mu>,\quad
C_2 \vert \mu > =\half \kappa_{\mu} \vert \mu>
\label{casimirs}
\ee
where $\vert \mu \vert =\sum_i \mu_i$ and $\half \kappa_{\mu}=\sum_{(i,j)\in \mu} i-j.$
As was discussed in \cite{ORV} operator $U_{-p}$ is responsible
for the change of slicing of the 3d partition.
More precisely, $U_{-p}$  maps
a 2d partition in a diagonal slice $x-y=t$ to the 2d partition
in the plane intersecting $z=0$ plane along the line 
$y=-t+(1+p)x$ and $x=0$ plane along the line $y=-t-pz.$ 
We illustrate the action of $U_{-p}$ in Figure \ref{fig10}.
We regard (\ref{conj}) as an indication that in the melting crystal picture
the initial lattice $\Lambda_{(p)} \in {\cal O}^+$ has vacancies in the y-z plane at the sites:
\be
 (0,y,z) \quad y < N_0-p z  
\label{newlattice}
\ee
Initial  lattice for the case of $p=-2$ is shown in Figure \ref{fig11}.

\section{D-branes in the crystal picture in the limit $g_s\to 0 $}
\begin{figure}
\begin{center}
\epsfig{file=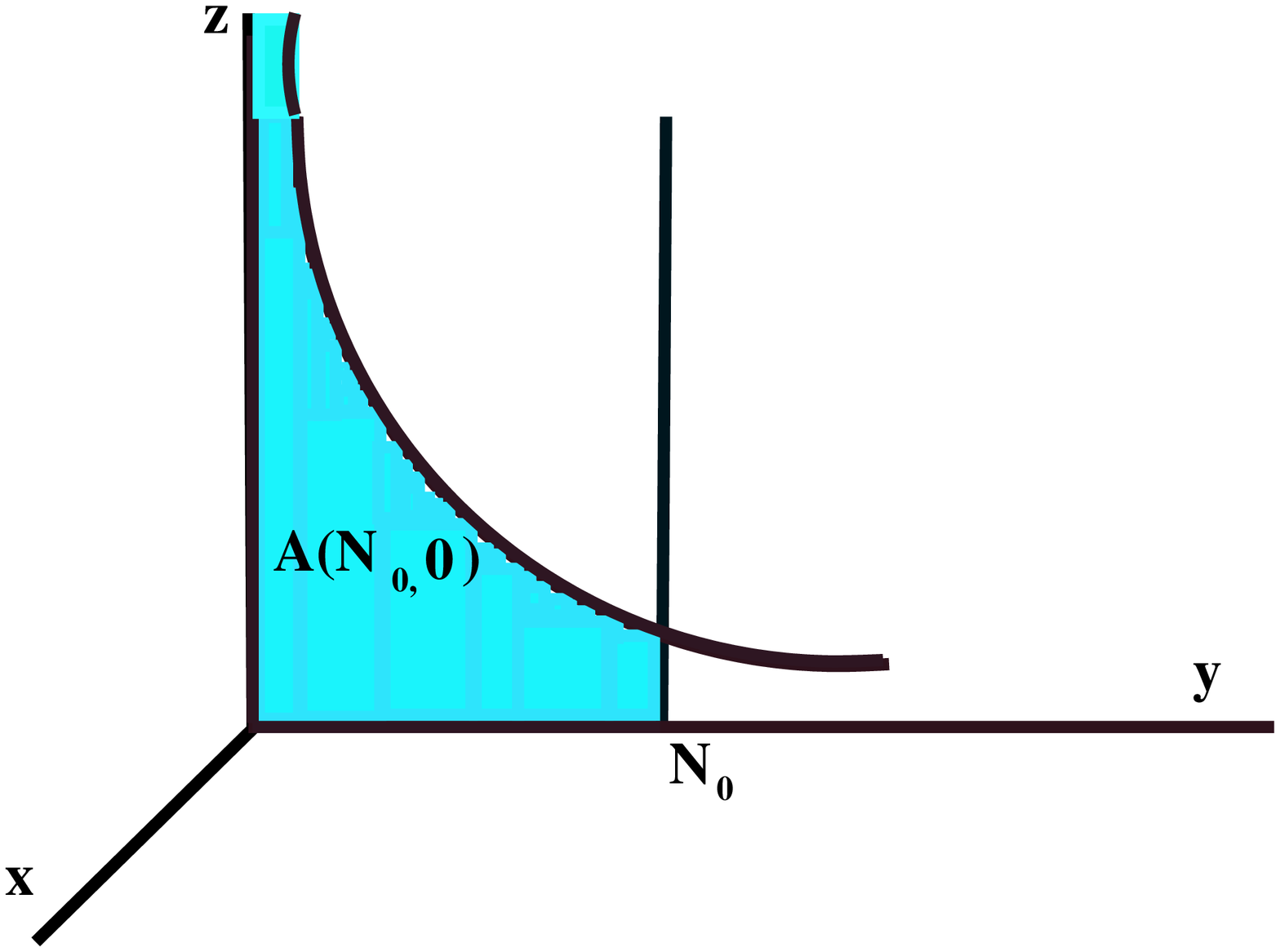,width=120mm}
\end{center}
\caption{ Area ${\cal A}(N_0,0).$ }
\label{fig12}
\end{figure}

In this section we demonstrate that our proposal for D-branes
in the crystal picture is consistent with string theory
in the semiclassical limit  $g_s \to 0.$
As shown in \cite{CK,OR} the lattice corner disappears in the limit
$g_s \to 0$ and smooth geometry, called the limiting shape, emerges.
The average configuration contains many  molten atoms and is given by the volume
above the 3d graph\footnote{ There is also  $x,y,z$ symmetric parametrization of the limiting shape
\cite{OR} }
\be
(x,y,z)=(R,U+R,V+R),\quad U=y-x,\quad  V=z-x,\quad R=R(U,V)
\label{shape}
\ee
 $R(U,V)$ is defined as
\be
R(U,V)=\frac{1}{4\pi^2}\int\int_{0}^{2\pi}log\vert F(U+i\a,V+i\b) \vert d\a d\b
\label{ronkin}
\ee
where
\be
F(u,v) =e^{-u}+e^{-v}+1
\label{mirror}
\ee
It is quite remarkable \cite{ORV} that the same function $F(u,v)$
enters the equation for the  Calabi Yau 3-fold mirror to $\C$
\be
Z_1Z_2=F(u,v)
\label{hyper}
\ee
Note that the 3d graph intersects $(y,z)$ plane in ${\cal O}^+$  over the curve
\be
e^{-y}+e^{-z}=1,\quad y\in [0,\infty]
\label{xzproject}
\ee
In the absence of D-branes the partition function behaves as:
\be
Z_{crystal}(q) \rightarrow exp\Bigl(-\frac{V_0}{g_s^2} \Bigr)
\label{semi}
\ee
where $V_0$ is the volume of the complement of the limiting shape (\ref{shape}). 
Now let us include  a zero-framing D-brane of the geometry (\ref{geom}).
We treat this as a small perturbation to the melting crystal and view
the additional D-brane defect in the background of the molten crystal.
We have proposed in section 3 that in this case
the melting starts from the lattice   in ${\cal O}^+$
with all the sites $(0,y,z), \quad y < N_0 g_s$ already vacant.  
This implies that the deleted region in the presence of the D-brane
$V_D$ differs from $V_0$ by the amount
\be
V_D=V_0-g_s{\cal A}(N_0,0)
\label{dif}
\ee
where ${\cal A}(N_0,0)$ stands for the area of the figure in yz-plane in ${\cal O}^+$
  below the curve (see Figure \ref{fig12} )
\be
e^{-y}+e^{-z}=1,\quad y\in [0,N_0 g_s]
\label{xzprojii}
\ee
But it is exactly ${\cal A}(N_0,0)$ that was shown in \cite{AKV}
to be the genus zero open string free energy in the A-model 
 on $\C$ in the presence of  a zero-framing D-brane.
This proves that our proposal in section 3 for the zero-framing D-branes in crystal melting
picture is consistent with the string theory results.

\begin{figure}
\begin{center}
\epsfig{file=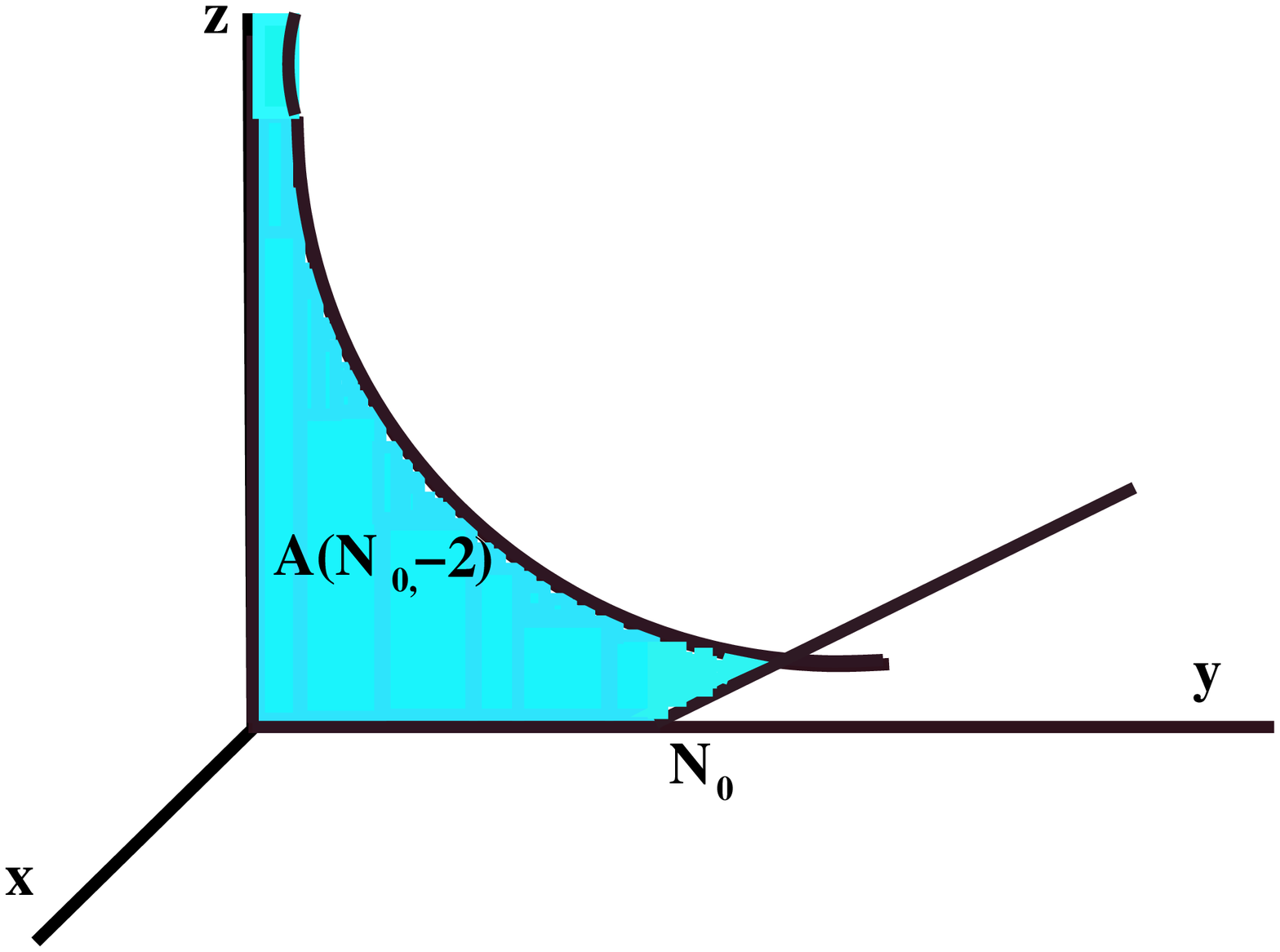,width=120mm}
\end{center}
\caption{ Area ${\cal A}(N_0,-2).$ }
\label{fig13}
\end{figure}

From \cite{AKV} we also get evidence for the crystal interpretation of framing
proposed in section 4. As explained in \cite{AKV} the genus zero open string free energy
 in the presence of  a  D-brane with framing $p$ is given by the area ${\cal A}(N_0,p)$
of the figure in yz-plane in ${\cal O}^+$
  below the curve (see Figure \ref{fig13} )
\be
e^{-y}+e^{-z}=1,\quad y\in [0,N_0 g_s-pz]
\label{xzprojiii}
\ee

This is consistent with our proposal in section 4 that 
the melting starts from the lattice   in ${\cal O}^+$
with all the sites $(0,y,z), \quad y < N_0 g_s-pz$ already vacant.  
Indeed, in this case  the deleted region  would have the volume
\be
V_{D,p}=V_0-g_s{\cal A}(N_0,p)
\label{difp}
\ee

{\bf Acknowledgments}

We would like to thank M. Aganagic for many valuable discussions.  In addition
we have benefited from discussions with  A. Okounkov and N. Reshetikhin.
 This research was supported in part
by NSF grants PHY-0244821 and DMS-0244464.


\end{document}